# Paramagnetic Spin Hall Magnetoresistance

Koichi Oyanagi[1,2*†], Juan M. Gomez-Perez[3*], Xian-Peng Zhang[4,5], Takashi Kikkawa[1,6,7], Yao Chen[1,6], Edurne Sagasta[3], Andrey Chuvilin[3,8], Luis E. Hueso[3,8], Vitaly N. Golovach[4,5,8], F. Sebastian Bergeret[4,5], Fèlix Casanova[3,8†], and Eiji Saitoh[1,6,7,9,10]

1. Institute for Materials Research, Tohoku University, Sendai 980-8577, Japan

2. Faculty of Science and Engineering, Iwate University, Morioka 020-8551, Japan

3. CIC nanoGUNE, 20018 Donostia-San Sebastian, Basque Country, Spain

4. Donostia International Physics Center, 20018 Donostia-San Sebastian, Basque Country, Spain

5. Centro de Fisica de Materiales (CFM-MPC), Centro Mixto CSIC-UPV/EHU, 20018 Donostia-San Sebastian, Basque Country, Spain

6. WPI Advanced Institute for Materials Research, Tohoku University, Sendai 980-8577, Japan

7. Department of Applied Physics, The University of Tokyo, Tokyo 113-8656, Japan

8. IKERBASQUE, Basque Foundation for Science, 48013 Bilbao, Basque Country, Spain

9. Center for Spintronics Research Network, Tohoku University, Sendai 980-8577, Japan

10. Advanced Science Research Center, Japan Atomic Energy Agency, Tokai 319-1195, Japan

*These authors contributed equally to this work.

†Correspondence and requests for materials should be addressed to K.O. (email: k.0yanagi444@gmail.com) or F.C. (email: f.casanova@nanogune.eu)

We report the observation of the spin Hall magnetoresistance (SMR) in a paramagnetic insulator. By measuring the transverse resistance in a Pt/Gd$_3$Ga$_5$O$_{12}$ (GGG) system at low temperatures, paramagnetic SMR is found to appear with an intensity that increases with the magnetic field aligning GGG's spins. The observed effect is well supported by a microscopic SMR theory, which provides the parameters governing the spin transport at the interface. Our findings clarify the mechanism of spin exchange at a Pt/GGG interface, and demonstrate tunable spin-transfer torque through the field-induced magnetization of GGG. In this regard, paramagnetic insulators offer a key property for future spintronic devices.

Spintronics [1,2] aims to add new functionalities to the conventional electronics using interconversion of spin angular momentum between different carriers in solids. Especially, the spin exchange between conduction-electron spins in a normal metal (NM) and magnetization, **M**, in a ferromagnet (FM) is a central topic to the branch of spintronics trying to manipulate **M** for developing new types of magnetic memory devices [3,4]. When spin angular momentum is transferred into a FM through a NM/FM interface [Fig. 1(a)], it modifies the transverse dynamics of **M** by exerting two types of torque, known as spin-transfer (damping-like) torque [5,6] and field-like torque [7], while it hardly couples to the longitudinal component. This is because the magnetic susceptibility in spin order, such as FM, is anisotropic due to the broken rotational symmetry reflecting spontaneous **M**; the magnetic susceptibility is large (small) along the transverse (longitudinal) direction, resulting in anisotropy into the spin injection.

The efficiency of the transverse spin injection has been characterized by the spin-mixing conductance $G_{\uparrow\downarrow}$ [8,9]. Its evaluation is of crucial importance in spintronics as $G_{\uparrow\downarrow}$ governs the device performance [10]. To this end, the spin Hall magnetoresistance (SMR) [11-21] can be a powerful tool. SMR is a resistance modulation effect in NM caused by a spin-current flow in NM and spin injection across a NM/FM interface. So far, SMR has been detected in NMs with various ordered (ferri-, ferro-, and antiferro-) magnets [11-21], which quantified $G_{\uparrow\downarrow}$ in the magnets. SMR has also been reported in some paramagnetic systems [22-25], but the mechanism of the effects has not been elucidated.

In this Letter, we demonstrate spin Hall magnetoresistance in a paramagnetic insulator (PI) $Gd_3Ga_5O_{12}$ (GGG), with a NM (Pt) contact. Unlike ordered magnets, a paramagnet has no spontaneous magnetization and shows huge longitudinal susceptibility. At the interface, conduction-electron spins in the NM couple not only to the transverse component (spin-transfer and field-like torque) but also to the longitudinal component of spins in PI through the interfacial spin-flip process [Fig. 1(b)], whose efficiency is characterized by the effective spin conductance (or spin-sink conductance) $G_s$ [9,26-29]; both $G_{\uparrow\downarrow}$ and $G_s$ are crucial for spin exchange at NM/PI interfaces. First, we show evidence of the paramagnetic SMR in Pt/GGG through transverse resistivity measurements. By combining experimental and theoretical results, we then evaluate $G_{\uparrow\downarrow}$ and $G_s$, and demonstrate that these spin conductances are controllable with external magnetic fields $B$. Such controllability in paramagnets is distinct from SMR in ordered magnets, highlighting the novelty of the paramagnetic SMR.

The sample consists of a Pt Hall bar (thickness $d$ = 5 nm, width $w$ = 100 μm, and length $l$ = 800 μm) on a single-crystalline GGG (111) slab. We measured longitudinal and transverse resistivity $\rho_L = wdR_L/l$ and $\rho_T = dR_T$, where $R_L = V_L/J_c$ ($R_T = V_T/J_c$,) is the Pt longitudinal (transverse) resistance [15] (see Appendix A) by applying current $J_c$ of typical amplitude of 200 μA using a DC reversal method [30] with applying $B$ up to 9 T.

Figure 1(c) shows the temperature ($T$) dependence of $M$ of the GGG slab measured by a vibrating sample magnetometer. The $M$-$T$ curve follows the Curie-Weiss law down to 2 K with a very low Curie-Weiss temperature $\Theta_{CW}$ = -2 K. $M$ arises from $Gd^{3+}$ spins ($S$ = 7/2), which are coupled via a weak exchange interaction [31] of 0.1 K. Because of the half-filled 4$f$-shell in $Gd^{3+}$, the orbital angular moment is zero, leading to the very small magnetic anisotropy of 0.04 K [31], which makes GGG an ideal paramagnetic system.

We have investigated paramagnetic SMR in a Pt/GGG junction system shown in Fig 1(d). SMR originates from a combination of the direct and inverse spin Hall effects (SHE and ISHE) [32-34]. When the charge current $\mathbf{J_c}$ is applied to the Pt layer, SHE creates a conduction-electron spin current, $\mathbf{J_s}$, with the spin polarization $\boldsymbol{\sigma}$ flowing along the $\boldsymbol{\sigma} \times \mathbf{J_c}$ direction. When the spin current $\mathbf{J_s}$ reaches the interface, it is reflected back into the Pt layer and again converted into a charge current via ISHE, causing the modulation of the Pt resistivity $\rho_{Pt}$. We can tune the reflected spin current and thereby $\rho_{Pt}$ by the field-induced magnetization $\mathbf{M} \sim \langle \mathbf{S}_\parallel \rangle$ of GGG. At the Pt/GGG interface, conduction-electron spins in Pt interact with the paramagnetic spins $\mathbf{S}$ in GGG via the interface exchange interaction, that exerts torque on $\mathbf{S}$. This torque is maximal (minimal) when $\boldsymbol{\sigma} \perp \langle \mathbf{S}_\parallel \rangle$ ($\boldsymbol{\sigma} \| \langle \mathbf{S}_\parallel \rangle$), where the intensity of the reflected spin current and the resultant ISHE are suppressed (enhanced). Therefore, $\rho_{Pt}$ becomes higher for $\boldsymbol{\sigma} \perp \langle \mathbf{S}_\parallel \rangle$ than for $\boldsymbol{\sigma} \| \langle \mathbf{S}_\parallel \rangle$. Besides, the effective magnetic field due to the interface exchange interaction affects the motion of conduction electrons in the Pt layer and gives an additional Hall component, referring to the spin Hall anomalous Hall effect (SHAHE) [12,15,16].

SMR measurements at low temperatures have been very difficult so far. This is because, at low $T$ and high $B$, weak anti-localization (WAL) effects appear in magnetoresistance and mask SMR signals in a four-probe resistance method [35]. Indeed, we observed a clear WAL signal at $T$ = 2.5 K in $\Delta\rho_L(B_i) = [\rho_L(B_i) - \rho_L(0)]/\rho_L(0)$, where $i = x, y, z$ in Fig. 2(e) and discussed in Appendix B. To overcome the problem, we measured *transverse* resistivity of the Pt layer [see Fig. 2(d)], in which WAL does not appear even under $B$; the setup allows us to investigate magnetoresistance free from WAL at low $T$ and high $B$.

Figure 2(b) shows the field dependent magnetoresistance (FDMR) amplitude $\Delta\rho_T(B) = [\rho_T(B) - \rho_T(0)]/\rho_L(0)$ at 2 K with $B$ at $\alpha = 45°$, where the transverse SMR becomes the most prominent. The $B$-rotation angle $\alpha$ is defined in Fig. 2(d). We observed clear magnetoresistance at $\alpha = 45°$, in sharp contrast with the result at $\alpha = 0$. The observed magnetoresistance increases for $|B| < 5$ T, while it is saturated for $|B| > 5$ T. The $B$ range, at which $\Delta\rho_T(B)$ is saturated is similar to that of $M$ [see Fig. 2(a)], suggesting the field-induced paramagnetism plays a dominant role.

SMR can be discussed in terms of the $\alpha$ dependence, which is phenomenologically given by $\cos(\alpha)\sin(\alpha)$ for the transverse component [11,12]. Figure 3(b) shows the angular dependent magnetoresistance (ADMR) of $\Delta\rho_T$ at 2 K by changing $\alpha$ at $|B| = 3.5$ T [see Fig. 3(d)]. $\Delta\rho_T(\alpha)$ shows a clear $\cos(\alpha)\sin(\alpha)$ feature, consistent with the transverse SMR scenario. Figure 3(f) shows the ADMR results at several $B$ values, which are well described by $S_{SMR}^{ADMR}(B)\cos(\alpha)\sin(\alpha)$ (except for $B = 0$). $S_{SMR}^{ADMR}(B)$ is plotted in Fig. 3(a) (purple circles), showing good agreement with the FDMR result (blue solid line), $S_{\text{SMR}}^{\text{FDMR}} = \Delta\rho_T(45°) - \Delta\rho_T(135°)$. The Hanle magnetoresistance (HMR) may cause a similar signal [36]. Figure 2(f) shows the $B$ dependence of HMR, $L_{\text{HMR}}(B) = [\rho_L(B_x) - \rho_L(B_y)]/\rho_L(0)$ at 300 K. We confirmed no meaningful signal from HMR at 300 K in our sample. The same claim can be made at 2 K because HMR weakly depends on $T$ [36,37] (see details in Appendix C). We thus conclude that the observed FDMR and ADMR are the experimental signatures of the paramagnetic SMR.

We found that the paramagnetic SMR manifests itself even in longitudinal resistivity measurements. Figure 3(c) shows the ADMR amplitude $\Delta\rho_L(\alpha) = [\rho_L(\alpha) - \rho_L(90°)]/\rho_L(0)$ at 2 K and $|B| = 3.5$ T [see Fig. 3(e)]. $\Delta\rho_L(\alpha) = [\rho_L(\alpha) - \rho_L(90°)]/\rho_L(0)$ is described by $L_{\text{SMR}}^{\text{ADMR}}\cos^2(\alpha)$, consistent with the expected behavior of SMR, i.e., the higher (lower) resistivity for $\mathbf{J_c}\|\mathbf{B}$ ($\mathbf{J_c}\perp\mathbf{B}$). Except for $B = 0$, similar $\cos^2(\alpha)$ dependence was confirmed at several $B$ values [Fig. 3(g)], and $L_{\text{SMR}}^{\text{ADMR}}(B)$ matches $S_{\text{SMR}}(B)$ [Fig. 3(a)]. Therefore, even from the longitudinal FDMR results, we successfully discerned the paramagnetic SMR from the WAL background signals (see Appendix D for further discussion).

We briefly argue the $\alpha$, $\beta$, and $\gamma$ dependence of $\Delta\rho_L$ in Figs. 4(a) and (b). In contrast to $\Delta\rho_L(\alpha)$, large WAL signals appear in $\Delta\rho_L(\beta)$ and $\Delta\rho_L(\gamma)$, which deviate from a $\cos^2$ dependence. The phenomenology of SMR and WAL explains the results as $\Delta\rho_L(\alpha)$: SMR only, $\Delta\rho_L(\beta)$: SMR + WAL, and $\Delta\rho_L(\gamma)$: WAL only. We indeed found $\Delta\rho_L(\beta) - \Delta\rho_L(\gamma) \sim \Delta\rho_L(\alpha)$. Therefore, all the ADMR results are ascribable to WAL and the paramagnetic SMR.

Figure 4(c) shows $S_{\text{SMR}}^{\text{ADMR}}(T)$ at $|B|$ = 3.5 T. $S_{\text{SMR}}$ shows the maximum value at 2 K and monotonically decreases with increasing $T$, resembling $M$-$T$ of GGG [the inset to Fig. 4(c)]. The results again show the field-induced $M$ is important to generate $S_{\text{SMR}}$, consistent with the paramagnetic SMR scenario.

Figure 2(c) shows $\Delta\rho_{\text{T}}(B)$ measured with applying $B \| z$ [sketch in the inset to Fig. 2(c)]. After subtracting the $B$-linear ordinary Hall effect (OHE) component, we found a small $B$-nonlinear signal $S_{\text{SHAHE}}$ for $|B|$ < 5 T at 2 K. For positive (negative) $B$, a positive (negative) signal appears; this $B$-odd dependence is characteristic of SHAHE [12,15,16,18]. With increasing $B$, $S_{\text{SHAHE}}$ increases and is saturated at around 5 T, concomitant with the saturation of $M$ in GGG [Fig. 2(a)]. We confirmed the higher-order SHAHE [16] is negligible (see Appendix E).

We apply a microscopic SMR theory [26] valid for NM/PI with the *B-dependent* magnetization instead of the phenomenological SMR theory [12] for NM/FM with the spontaneous *B-independent* magnetization, leaving the $B$ dependence of SMR unexplained. We describe the spin current $\mathbf{J_s}$ at the NM/PI interface resulting from the interfacial exchange interaction by using the boundary condition [9,26,28] written as

$$-e\mathbf{J}_s = G_r \mathbf{n} \times (\mathbf{n} \times \boldsymbol{\mu}_s) + G_i \mathbf{n} \times \boldsymbol{\mu}_s + G_s \boldsymbol{\mu}_s, \qquad (1)$$

where $e$ is the elementary charge, $\mathbf{n}$ the unit vector of $\mathbf{B}$, $\boldsymbol{\mu}_s$ the spin accumulation in the NM side, $G_{\uparrow\downarrow} = G_r + iG_i$ the spin-mixing conductance, and $G_s$ the effective spin conductance. The first and second terms in the right-hand side of Eq. (1) correspond to the spin-transfer and field-like torque, respectively, and the third indicates the spin-flip (electron-magnon) scattering, which accounts for the magnon-related unidirectional SMR [38-40]. We calculate the spin conductances in Eq. (1) for the NM/PI interface as

$$G_r(B) = A_1 \left\{ S(S+1) - \left[\coth(\xi/2) + \frac{\xi}{4\sinh^2(\xi/2)}\right] SB_S(S\xi) \right\}, \qquad (2)$$

$$G_i(B) = A_2 SB_S(S\xi), \qquad (3)$$

$$G_s(B) = -A_1 \frac{\xi}{2\sinh^2(\xi/2)} SB_S(S\xi), \qquad (4)$$

where $B_S(x)$ is a Brillouin function of spin-$S$ as a function of $x$, $\xi(B) = C_1 B[T/(T - \Theta_{CW}^{\text{eff}})]$, $\Theta_{CW}^{\text{eff}}$ the effective Curie-Weiss temperature, which contains $\Theta_{CW}$, $S = 7/2$ the electron spin of a $Gd^{3+}$ ion, and $C_1$ a numerical constant. $A_{1,2}$ are fitting parameters, which contain the interface spin density $n_{PI}$ and dimensionless interfacial $s$-$f$ exchange interaction $J_{\text{int}}$. Finally, the magnetoresistance as a function of $B$ is given by:

$$S_{\text{SMR}}(B) = D_1 \{\mathcal{R}(G_s) - \text{Re}[\mathcal{R}(G_s - G_{\uparrow\downarrow})]\}, \qquad (5)$$

$$S_{\text{SHAHE}}(B) = D_1 \text{Im}[\mathcal{R}(G_s - G_{\uparrow\downarrow})], \qquad (6)$$

where $\mathcal{R}(x) = (1 - D_2 x)/(1 - D_3 x)$, and $D_{1,2,3}$ are known numerical constants. We refer Appendices F to I for theoretical details. We obtained the best fits using Eqs. (5) and (6) simultaneously as shown in Fig. 5(a) with the values of $n_{PI} = 6.94 \times 10^{16}$ Gd atom/m$^2$, $\Theta_{CW} = -1.27$ K, and $J_{\text{int}} = -0.13$. Although a direct fit to $S_{\text{SMR}}(T)$ is not possible by simply considering $M(T)$, our model fully explains $S_{\text{SMR}}(B,T)$, in which $T$ dependences of the spin-transport parameters of the Pt film and effects from a paramagnetic subsystem are also taken into account (see Appendix J for further discussion).

Figure 5(b) shows $G_r(B)$, $G_i(B)$, and $G_s(B)$ with the estimated parameter values. At zero magnetic field, $G_r$ and $G_i$ vanish, while $|G_s|$ takes the maximum value of $8.7 \times 10^{12}$ S/m$^2$. By increasing $B$, both $G_r$ and $|G_i|$ monotonically increases, but $|G_i|$ increases more rapidly than $G_r$, and $G_r$ ($|G_i|$) approaches the value of $1.0 \times 10^{13}$ S/m$^2$ ($7.4 \times 10^{12}$ S/m$^2$) at around 5 T (3 T). On the other hand, $|G_s|$ monotonically decreases with $B$ and approaches zero.

The $B$-dependent spin transport at the interface is a unique feature of paramagnets, in sharp contrast to FM where $G_{\uparrow\downarrow}$ is almost independent of $B$. At the NM/PI interface, all the torque is cancelled out with the randomized spin ($\langle S_\parallel \rangle = 0$) at $B = 0$, resulting in $G_r = G_i = 0$. When the PI acquires a net magnetization with applying $B$, a positive $G_r$ and negative $G_i$ appear; the latter implies antiferromanetic $s$-$f$ interaction at the interface. On the other hand, $|G_s|$ decreases with $B$ due to the Zeeman gap ($\propto g\mu_B B$, where $g$ is the $g$-factor and $\mu_B$ the Bohr magneton). At small $B$, the localized spin can be easily flipped by spin and energy transfer between the conduction electron and localized spin. By applying $B$, the degeneracy of the paramagnetic spin is lifted to split into different energy levels by the Zeeman effect. Because the energy scale of the SHE-induced spin-flip scattering is governed by $k_B T$, where $k_B$ is the Boltzmann constant, at 2 K it can be suppressed by increasing $B$ (9 T for electrons corresponds to the energy scale of 25 K), leading to the reduction of $G_s$.

Our results clarify the mechanism of SMR and SHAHE in paramagnets. By comparing $S_{SMR}$, $S_{SHAHE}$, $G_r$, and $|G_i|$, we found $S_{SMR}(B) \propto G_r(B)$ and $S_{SHAHE}(B) \propto |G_i(B)|$ in Figs. 5(e) and (f), respectively. Because $G_r$ and $|G_i|$ represent the efficiencies of the spin-transfer and field-like torque, respectively [Figs. 5(c) and (d)], the agreement indicates that the spin-transfer (field-like) torque is the mechanism of SMR (SHAHE) in Pt/GGG. Furthermore, the

agreement between the experiment and theory clarifies that SMR is attributed to the ensemble of paramagnetic moments, consistent with the scenario in other magnetic ordered systems. This contrasts with the conclusions of Ref. 25, in which the MR observed in non-crystalline paramagnetic YIG/Pt was attributed to the total magnetic moment. Our results thus unify the description of SMR in compensated ferrimagnets [41,42], antiferromagnets [19-21,43], ferromagnets [18], and paramagnets, resolving the longstanding controversy for the origin of SMR.

Finally, we discuss the interfacial parameters $J_{int}$, $n_{PI}$, and $\Theta_{CW}$. We obtained a negative interfacial exchange interaction of about -2 meV (see Appendix I). This value has the same sign and order of magnitude as the one found in the Pt/EuS interface [18,44], -3~-4 meV, indicating the *s-f* exchange coupling is antiferromagnetic in both systems. On the other hand, a negative $G_i$ was found in W/EuO [45], corresponding to a positive (ferromagnetic) *s-f* exchange interaction. The sign of the exchange interaction in metallic compounds with rare-earth ions depends on the electron structure of the host metal and the type of the rare-earth ions [46], and so may do the interfacial exchange interaction. The estimated Gd atom density corresponds to only 1% of the bulk value for GGG. The depletion of Gd atoms at the interface is consistent with the smaller $\Theta_{CW}$ of -1.27 K than the bulk value of -2 K, indicating the decrease of the exchange interaction among Gd atoms at the interface. The feature may be attributed to possible damage of the GGG surface crystallization during the Pt sputtering (see the TEM images in Appendix J).

In summary, we demonstrate the paramagnetic SMR in a Pt film on GGG at 2 K. The SMR is induced with applying magnetic fields, and saturated above several tesla when all localized spins are aligned. The observed correlation between SMR/SHAHE and magnetization indicates that the field-induced magnetization plays a significant role in the spin transport at the Pt/GGG interface. Our microscopic theory well explains the SMR signals as a function of magnetic fields and quantifies the microscopic spin exchange parameters at the Pt/GGG interface. Our results indicate that the magnetoresistance measurements allow us to investigate spin transport at interfaces, essential for accelerating insulator-based spintronics.

**Acknowledgement**

We thank G. E. W. Bauer, B. J. van Wees, S. T. B. Goennenwein, S. Takahashi, and K. Sato for discussions. This work in Japan is a part of the research program of ERATO "Spin Quantum Rectification Project" (No. JPMJER1402) and CREST (Nos. JPMJCR20C1 and JPMJCR20T2) from JST, the Grant-in-Aid for Scientific Research on Innovative Area "Nano Spin Conversion Science" (No. JP26103005), the Grant-in-Aid for Scientific Research (S) (No. JP19H05600), Grant-in-Aid for Scientific Research (B) (No. JP20H02599), Grant-in-Aid for Research Activity start-up (No. JP20K22476), and Grant-in-Aid for Early-Career Scientists (No. 21K14519) from JSPS KAKENHI, and JSPS Core-to-Core program "the International Research Center for New-Concept Spintronics Devices", World Premier International Research Center Initiative (WPI) from MEXT, Japan. The work in Spain is supported by the Spanish MICINN under the Maria de Maeztu Units of Excellence Programme (MDM-2016-0618), and projects No. MAT2015-65159-R, FIS2017-82804-P, and RTI2018-094861-B-100. The work of F.S.B. is partially funded by EU's Horizon 2020 research and innovation program under Grant Agreement No. 800923 (SUPERTED). K.O. and Y.C. acknowledges support from GP-Spin at Tohoku University. Y.C. is supported by JSPS through a research fellowship for young scientists (No. JP18J21304). J.M.G.-P. and E. Sag. thank the Spanish MICINN and MECD for a Ph.D. fellowship (Grants No. BES-2016-077301 and FPU14/03102, respectively).

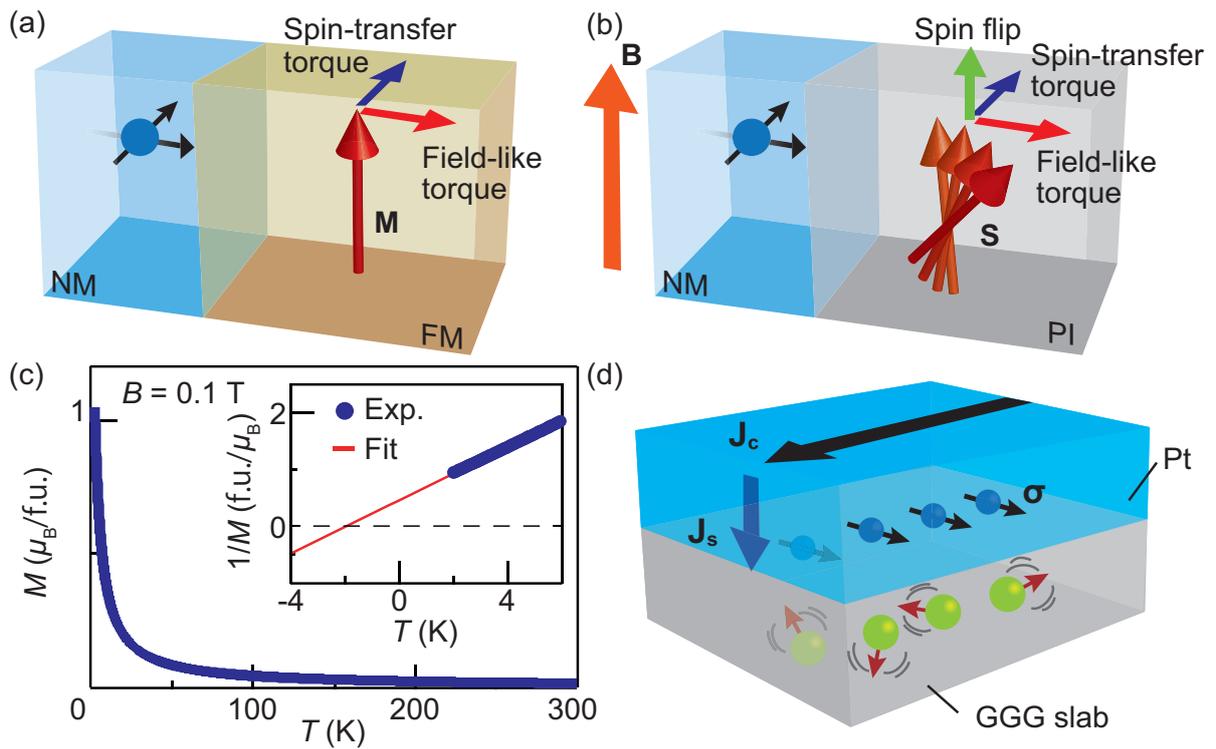

FIG. 1. (a) NM/FM and (b) NM/PI interface with spin exchange. The blue, red, and green arrows represent the directions of angular momentum related to the spin-transfer torque, field-like torque, and spin-flip process, respectively. (c) $M(T)$ of GGG. The inset shows $1/M$ (blue circles) and a linear fit (red solid line). (d) The Pt/GGG interface. $\mathbf{J_s}$ with $\boldsymbol{\sigma}$ is generated in Pt by SHE with the application of $\mathbf{J_c}$.

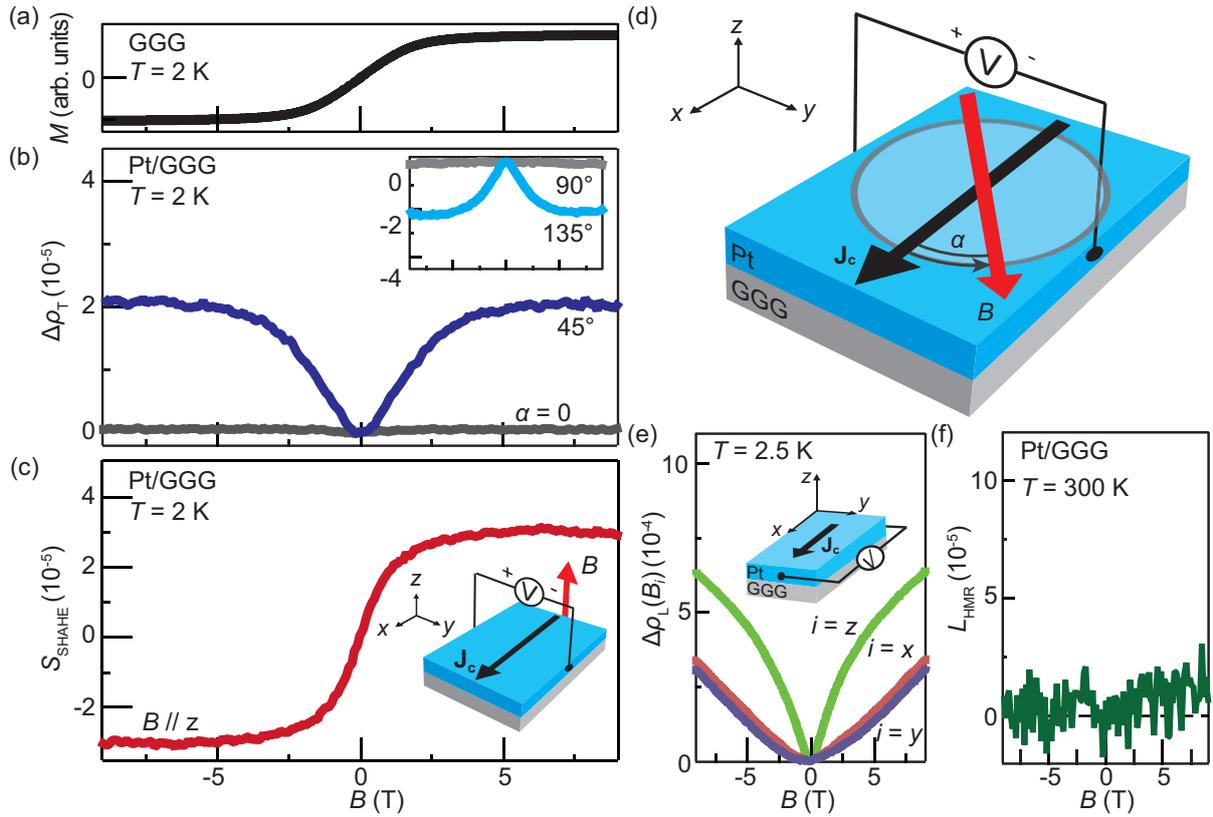

FIG. 2. (a) $M(B)$ of GGG at 2 K. (b) $\Delta\rho_T(B)$ at 2 K. The deep blue (grey) curve shows $\Delta\rho_T$ with $B$ at $\alpha = 45°$ (0°). The light blue (grey) curve in the inset shows $\Delta\rho_T$ with $B$ at $\alpha = 135°$ (90°). (c) $S_{SHAHE}(B)$ with $B||z$ after subtracting the OHE component. The inset shows the measurement setup for SHAHE. (d) Measurement setup for SMR. (e) $\Delta\rho_L(B_i)$ in Pt/GGG at 2.5 K. The inset shows the measurement setup. (f) $L_{HMR}(B)$ at 300 K in Pt/GGG.

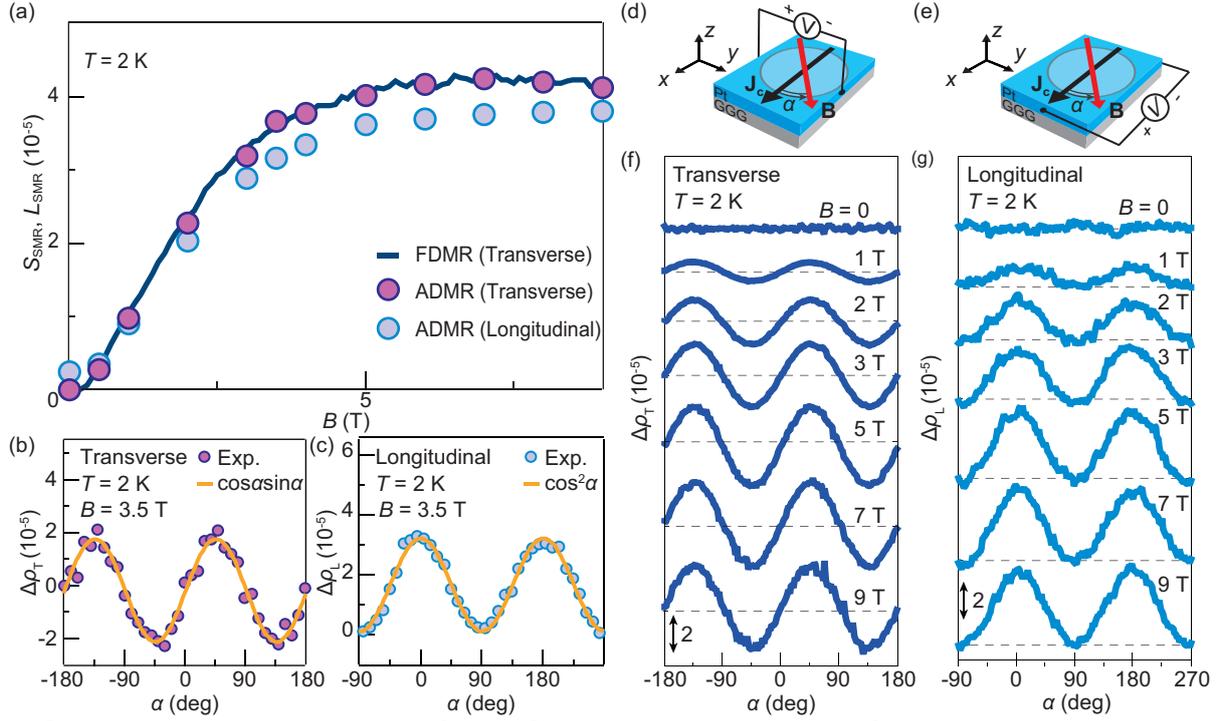

FIG. 3. (a) $B$ dependence of the SMR signals obtained from FDMR and ADMR measurements. The solid curve represents $S_{\text{SMR}}^{\text{FDMR}}$. The purple (blue) circles show $S_{\text{SMR}}^{\text{ADMR}}$ ($L_{\text{SMR}}^{\text{ADMR}}$) obtained via the fitting using $S_{\text{SMR}}^{\text{ADMR}}\cos(\alpha)\sin(\alpha)$ [$L_{\text{SMR}}^{\text{ADMR}}\cos^2(\alpha)$]. (b) $\Delta\rho_{\text{T}}(\alpha)$ and (c) $\Delta\rho_{\text{L}}(\alpha)$ at 2 K with rotating $|B|$ = 3.5 T. The orange solid curves in (b) and (c) are a $S_{\text{SMR}}^{\text{ADMR}}\cos(\alpha)\sin(\alpha)$ and $L_{\text{SMR}}^{\text{ADMR}}\cos^2(\alpha)$ fit, respectively. (d),(e) Schematic illustrations of (d) the transverse and (e) longitudinal ADMR measurement setup. (f) $\Delta\rho_{\text{T}}(\alpha)$ and (g) $\Delta\rho_{\text{L}}(\alpha)$ at 2K for various $B$.

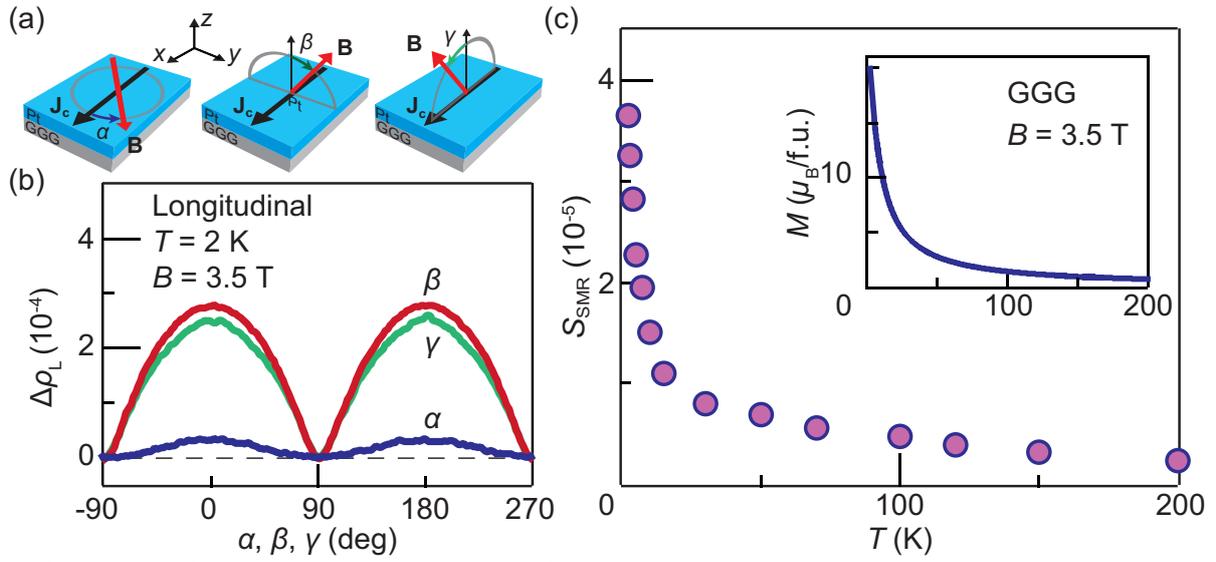

FIG. 4. (a) Schematic illustrations of the ADMR measurement setups. (b) $\Delta\rho_L(\alpha)$, $\Delta\rho_L(\beta)$, and $\Delta\rho_L(\gamma)$ data. (c) $S_{SMR}^{ADMR}(T)$ at 3.5 T. The inset shows the $M$-$T$ curve of GGG.

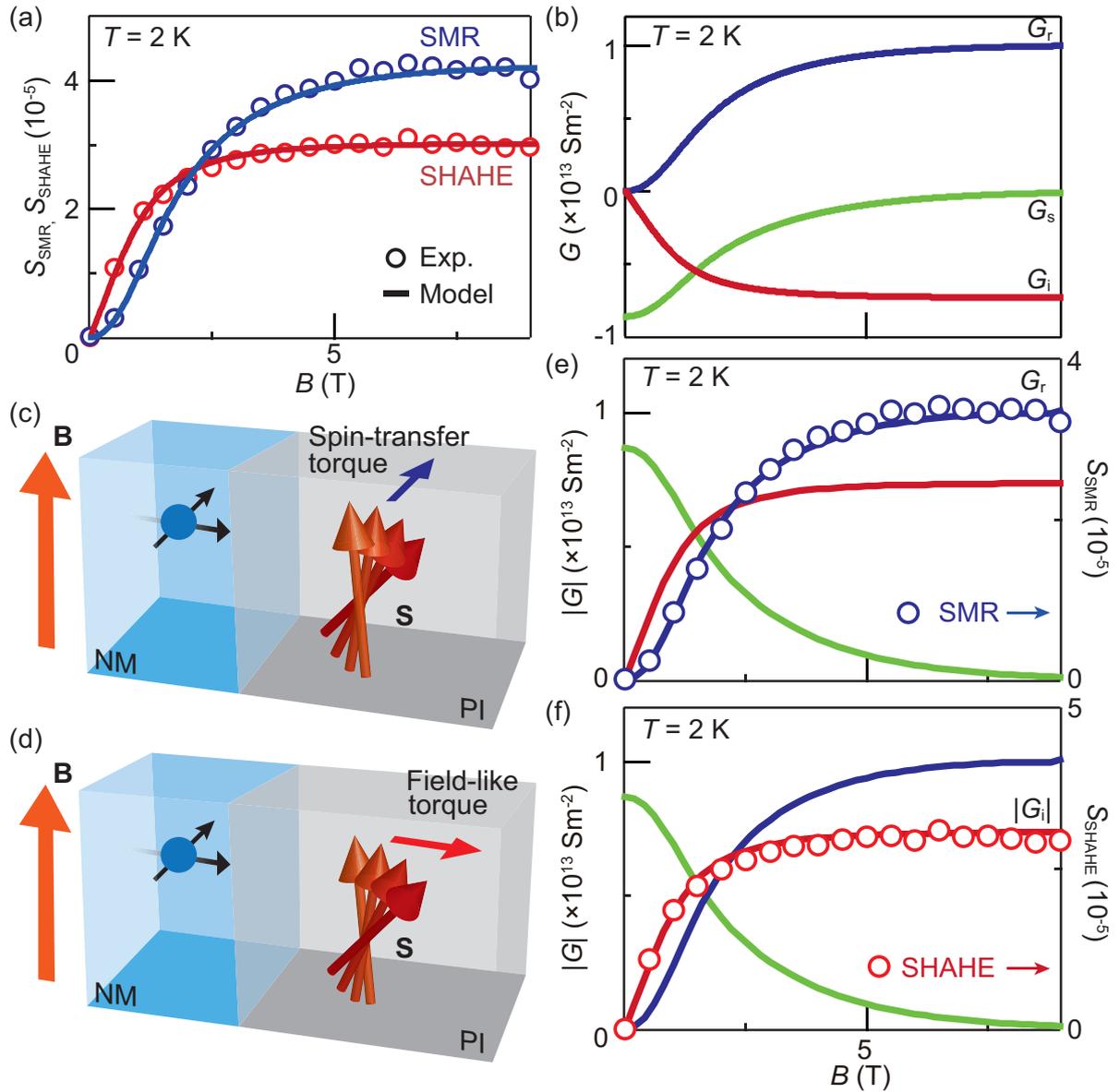

FIG. 5. (a) B dependence of SMR and SHAHE together with the fitting of Eqs. (5) and (6) at 2 K. (b) $B$ dependence of $G_r$, $G_i$, and $G_s$. (c) Spin-transfer and (d) field-like torque in a NM/PI system. (e),(f) The comparison (e) between $G_r$ and SMR, and (f) between $|G_i|$ and SHAHE.

**APPENDIX A: TEMPERATURE DEPENDENCE OF PT RESISTIVITY**

Figure 6 shows the temperature $T$ dependence of the resistivity $\rho_L(B = 0) = \rho_D$ of the Pt film on the GGG slab. We measured $\rho_L$ by the conventional four-probe method with applying charge current of 200 μA. Down to around 20 K, $\rho_L$ linearly decreases with decreasing $T$. Below 10 K, $\rho_L$ starts to increase, which is a signature of the weak anti-localization (WAL) effect [35].

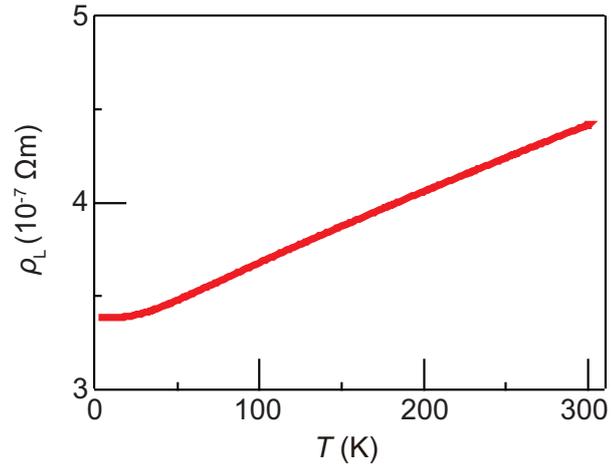

FIG. 6. $T$ dependence of $\rho_L$ of Pt. The resistivity $\rho_L$ is measured in a 5-nm thickness Pt Hall bar on the GGG slab.

## APPENDIX B: WEAK ANTI-LOCALIZATION IN PT FILM

We measured magnetoresistance in the longitudinal configuration [35] for $T < 50$ K to show the weak anti-localization (WAL) effects in the Pt film. Figure 7 shows the normalized longitudinal resistivity change $\Delta\rho_\text{L}(B_i) = [\rho_\text{L}(B_i) - \rho_\text{L}(B_i = 0)]/\rho_\text{L}(B_i = 0)$ as a function of $B$ in the $i = x$, $y$, and $z$ directions (see the inset to Fig. 7) at 2.5 K. $\Delta\rho_\text{L}(B_i)$ increases with increasing $B$ for all $i$, but their shapes differ with each other. $\Delta\rho_\text{L}(B_z)$ increases more rapidly, while $\Delta\rho_\text{L}(B_x)$ and $\Delta\rho_\text{L}(B_y)$ show similar trends. The largest value of $\Delta\rho_\text{L}(B_z)$ at 9 T is two times greater than those of $\Delta\rho_\text{L}(B_x)$ and $\Delta\rho_\text{L}(B_y)$. Note that the difference between $\Delta\rho_\text{L}(B_x)$ and $\Delta\rho_\text{L}(B_y)$, $\sim 4\times10^{-5}$ at 9 T, corresponds to the paramagnetic SMR (we discuss it in Appendix D).

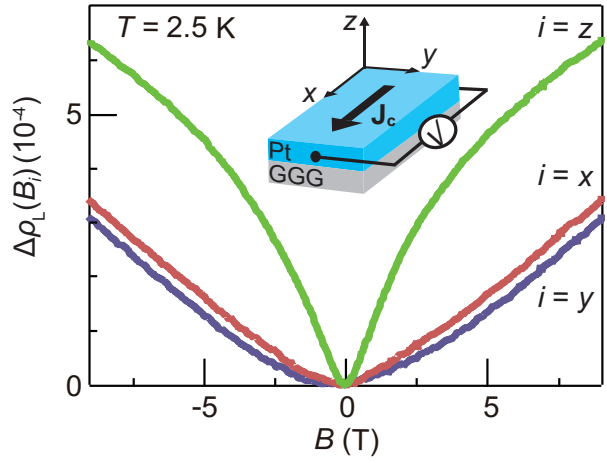

FIG. 7. $B$ dependence of $\Delta\rho_\text{L}(B_i)$ at 2.5 K. The red, blue and green curves indicate the $B$ dependence of $\Delta\rho_\text{L}(B_i)$ for $B$ in the $x$, $y$ and $z$ directions, respectively. The inset shows the sample with the applied current $\mathbf{J_c} \parallel \mathbf{x}$.

## APPPENDIX C: HANLE MAGNETORESISTANCE IN PT FILM

We here show that the Hanle magnetoresistance (HMR) [36] is negligibly small in our Pt/GGG sample. As shown in Refs. 36 and 37, the HMR may show *B*-direction dependence similar to the SMR [i.e., HMR appears (disappears) under *B* in the *x* and z (*y*) directions] and may show up in a wide temperature range from 300 K to 2 K, but weakly depends on *T*. To investigate the HMR in our sample, we measure the *B* dependence of the longitudinal resistivity at 300 K, where the HMR, if present in our sample, may only be detected, while the paramagnetic SMR is suppressed because of the negligibly small paramagnetic moment in GGG at such high temperatures. Figure 8 shows the *B* dependence of $L_{HMR} = [\rho_L(B_x) - \rho_L(B_y)]/\rho_L(B = 0)$ at $T = 300$ K. We found no magnetoresistance, indicating that the HMR is undetected in the present Pt/GGG system. We thus conclude that the HMR can be neglected in our sample in all the *T* range. The absence of the HMR in our Pt/GGG sample may be attributed to the Pt growth condition, the detail of which is discussed in Ref. 36.

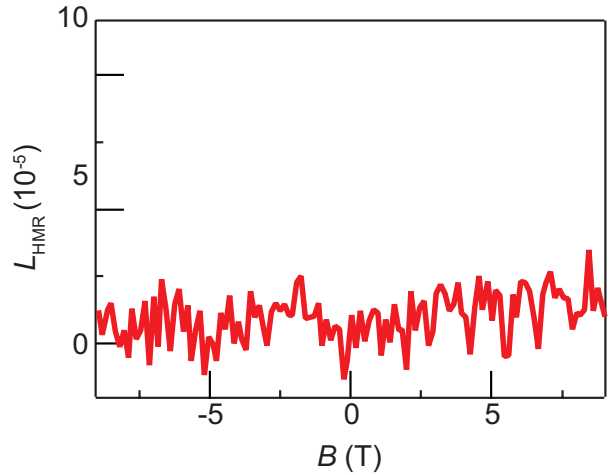

FIG. 8. *B* dependence of $L_{HMR}$ at 300 K.

## APPENDIX D: SMR IN LONGITUDINAL MEASUREMENTS IN PT/GGG

We show the paramagnetic SMR in the longitudinal configuration. We performed field dependent magnetoresistance (FDMR) measurements in the $x$ and $y$ directions and angular dependent magnetoresistance (ADMR) measurements in the $x$-$y$ plane [see Fig. 9(a)].

Figure 9(b) shows the $B$ dependence of $\Delta\rho_\text{L}(B_i) = [\rho_\text{L}(B_i) - \rho_\text{L}(B_i = 0)]/\rho_\text{L}(B = 0)$, where $i = x, y$ at 2.5 K. The overall behavior is ascribable to the WAL as discussed in Appendix B. We here address the small difference between $\Delta\rho_\text{L}(B_x)$ and $\Delta\rho_\text{L}(B_y)$, defined as $L_\text{SMR}^\text{FDMR}(B) = [\rho_\text{L}(B_x) - \rho_\text{L}(B_y)]/\rho_\text{L}(B = 0)$. We plot the $B$ dependence of $L_\text{SMR}^\text{FDMR}$ as a green curve in Fig. 9(d). $L_\text{SMR}^\text{FDMR}(B)$ gradually increases and approaches to $\sim 3\times10^{-5}$ at 5 T, similar to the $B$ dependence of $M$ in GGG and the FDMR result in the transverse configuration shown in Figs. 2(a) and 3(a).

Next, we investigate ADMR of the longitudinal resistivity in the $x$-$y$ plane. Figure 9(c) shows the $\Delta\rho_\text{L} = [\rho_\text{L}(\alpha) - \rho_\text{L}(\alpha = 90°)]/\rho_\text{L}(B = 0)$ as a function of $\alpha$ at $B = 3.5$ T and $T = 2.5$ K. $\Delta\rho_\text{T}$ shows a clear $L_\text{SMR}^\text{ADMR}\cos^2\alpha$ dependence, consistent with the behavior of the paramagnetic SMR. We extracted $L_\text{SMR}^\text{ADMR}$ at various $B$ and obtained the $B$ dependence of the ADMR result shown in Fig. 9(d). $L_\text{SMR}^\text{FDMR}$ and $L_\text{SMR}^\text{ADMR}$ agree well with each other. All experimental findings in the longitudinal measurements are consistent with the observed SMR in the transverse measurements, showing that the difference between $\Delta\rho_\text{L}(B_x)$ and $\Delta\rho_\text{L}(B_y)$ [Fig. 9(b)] can be attributed to the paramagnetic SMR.

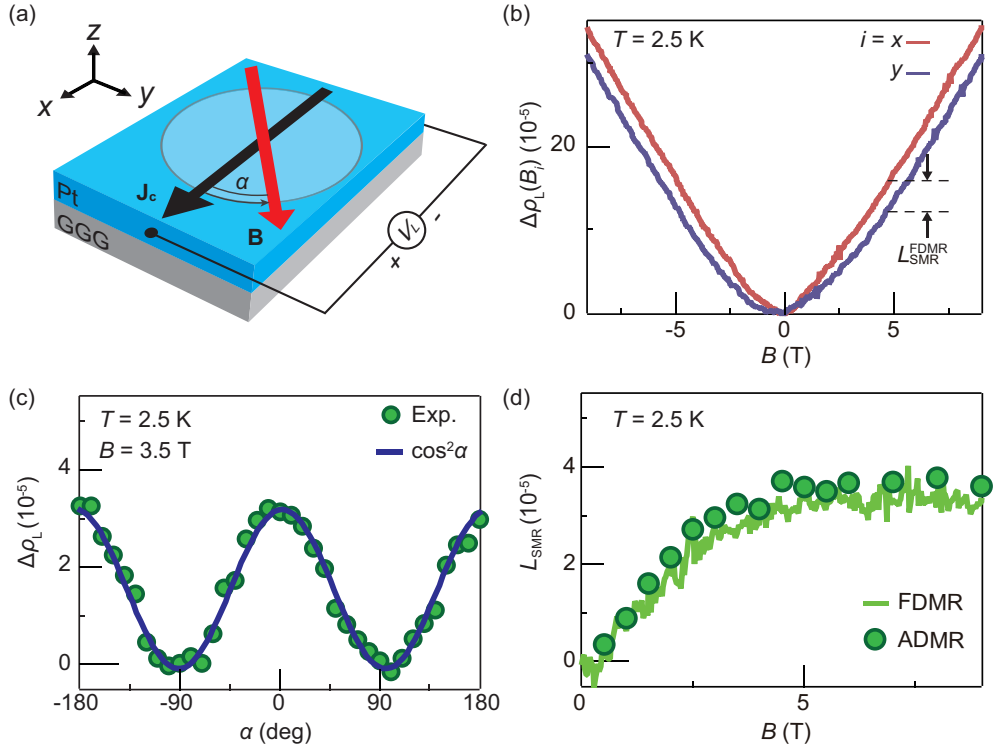

FIG. 9. (a) A schematic illustration of the longitudinal resistivity measurement. A charge current $\mathbf{J_c}$ is applied in the $x$ direction and the longitudinal voltage is measured. **B** indicates the spatial direction of the magnetic field, and the angle between $\mathbf{J_c}$ and $\mathbf{B}$ is defined as $\alpha$. (b) The field dependent magnetoresistance (FDMR) in the longitudinal measurement at 2.5 K. The red (blue) curve indicates $\Delta\rho_L$ under $B$ in the $x$ ($y$) direction. $L_{SMR}^{FDMR}(B)$ is defined as $L_{SMR}^{FDMR}(B) = [\rho_L(B_x) - \rho_L(B_y)]/\rho_L(B = 0)$. (c) The field angular dependent magnetoresistance (ADMR) $\Delta\rho_L$ in the $xy$-plane. The green circles show the $\alpha$ dependence of $\Delta\rho_L$ at 2.5 K. The blue curve indicates a $L_{SMR}^{ADMR}\cos^2\alpha$ fitting to the experimental result. (d) FDMR and ADMR results as a function of $B$ at 2.5 K. The green curve (circles) shows the FDMR (ADMR) result of $L_{SMR}$.

**APPENDIX E: SHAHE MEASUREMENT RESULT**

Figure 10(a) shows the $B$ dependence of the normalized Hall resistivity $\Delta\rho_\text{T}(B)$ at 2 K for Pt/GGG. $\Delta\rho_\text{T}$ increases linearly with increasing $B$ due to the ordinary Hall effect (OHE). We estimated the slope of the OHE $A_\text{OHE}$ as $-6.5\times10^{-5}$ (1/T), using a linear fitting to $\Delta\rho_\text{T}$ for the $|B| > 5$ T range and averaged the obtained slopes for $-9$ T $< B < -5$ T and 5 T $< B <$ 9 T. The obtained $A_\text{OHE}$ in the 5-nm-thick Pt film is almost the same as that obtained in the 30-nm-thick Pt film on GGG at low and high $T$ [47]. The absence of the Pt thickness and $T$ dependence of $A_\text{OHE}$ indicates that the contribution from the other Hall effects induced by the influence of the interface found in Pt/YIG is negligibly small [16,35,47]. After subtracting the OHE component, we found the spin Hall anomalous Hall effect (SHAHE) signal shown in Fig. 2(c).

Here we show that the higher-order contribution of SHAHE [16] is negligible small in our sample. Figure 10(b) shows the transverse ADMR at 2 K and selected magnetic fields. We clearly observe a $\cos\beta$ dependence of $\Delta\rho_\text{T}$ because of the OHE [see Fig. 10(a)]. To evaluate the higher-order contribution, we carried out the same analysis procedure used by Meyer et al. [16]: fitting a $A^{1st}\cos\beta + A^{3rd}\cos^3\beta$ function to the ADMR results shown in Fig. 10(b), where $A^{1st}(B) = S^{1st}_\text{SHAHE}(B) + A^{1st}_\text{OHE}\cdot B$, $S^{1st}_\text{SHAHE}(B)$ is the first-order SHAHE contribution as a function of $B$, $A^{1st}_\text{OHE}$ is the coefficient of the OHE, and $A^{3rd}(B) = S^{3rd}_\text{SHAHE}(B) + A^{3rd}_\text{OHE}\cdot B$ is the higher-order contribution of the same Hall terms. Figure 10(c) shows the $B$ dependence of $A^{1st}$ (blue solid circles) and $A^{3rd}$ (red solid circles). $A^{1st}$ and $A^{3rd}$ monotonically increase with increasing $B$ mainly due to the OHE, $A^{1st}_\text{OHE}$ and $A^{3rd}_\text{OHE}$, respectively. We estimate the slope of the OHE in the first- and high-order contribution as $A^{1st}_\text{OHE} = -6.3\times10^{-5}$ (1/T) and $A^{3rd}_\text{OHE} = -2.1\times10^{-6}$ (1/T) using the results at $B = 7$ T, and 9 T. $A^{1st}_\text{OHE} = -6.3\times10^{-5}$ (1/T) is similar to that estimated from the FDMR results [~ $-6.5\times10^{-5}$ (1/T)]. After subtracting the OHE, we show the $B$ dependence of $S^{1st}_\text{SHAHE}(B)$ and $S^{3rd}_\text{SHAHE}(B)$ in Fig. 10(d). We found the first-order SHAHE contribution [$S^{1st}_\text{SHAHE}(B) \sim 2.7\times10^{-5}$ at 9 T] is about 10 times larger than the higher-order one [$S^{3rd}_\text{SHAHE}(B) \sim 0.26\times10^{-5}$ at 9 T] in our system. Furthermore, $S^{1st}_\text{SHAHE}(B)$ from the ADMR results is consistent with $S_\text{SHAHE}(B)$ from the FDMR results [deep blue curve in Fig. 10(d), and shown in the manuscript]. Therefore, we can accurately obtain the amplitude of the first-order SHAHE from the FDMR result by subtracting the OHE component, justifying our analysis described as above.

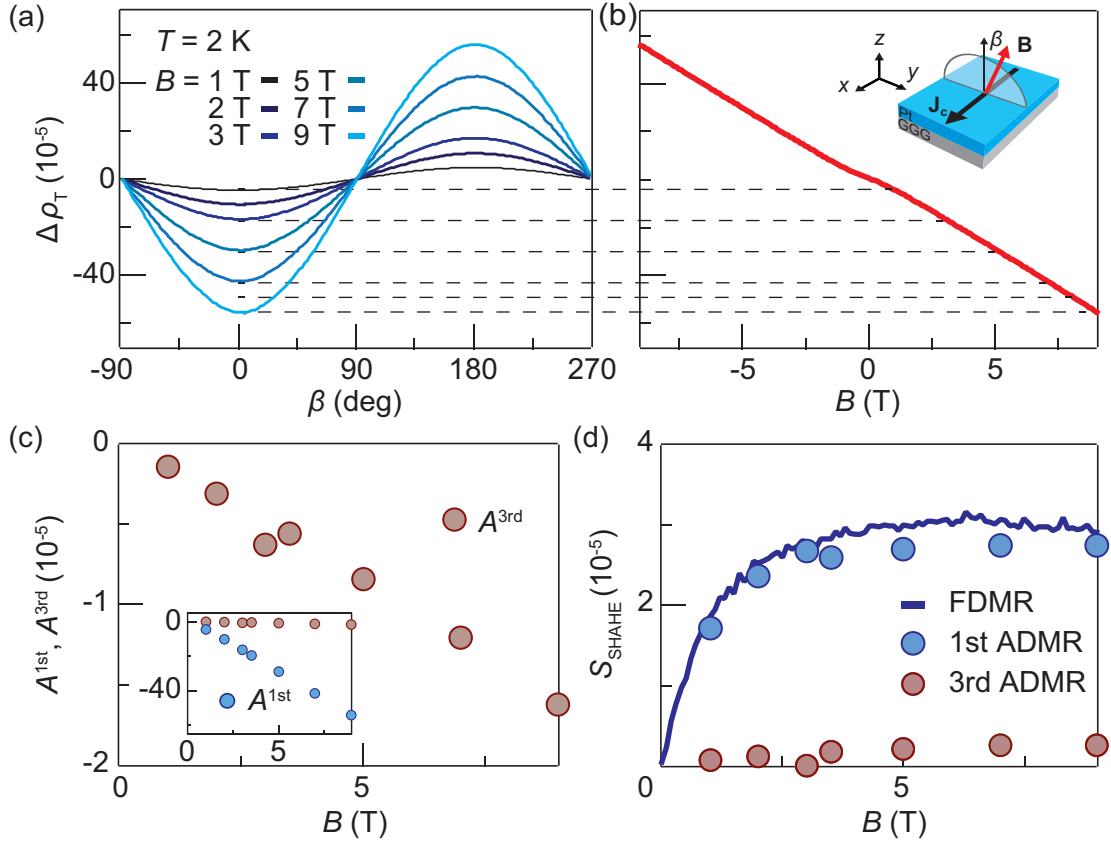

FIG. 10. (a) $B$ dependence of $\Delta\rho_T$ in the Hall measurement. $\mathbf{J_c}$, $\mathbf{B}$, and $\beta$ denote the spacial direction of the charge current and magnetic field, and relative angle, respectively. The Hall resistivity is measured at $\beta = 0$ and $T = 2$ K. (b) ADMR results of SHAHE at $T = 2$ K and selected $B$. (c) $B$ dependence of $A^{3rd}$. The inset shows the $B$ dependence of $A^{1st}$ and $A^{3rd}$. The amplitude of $A^{1st}$ and $A^{3rd}$ are obtained by fitting a $A^{1st}\cos\beta + A^{3rd}\cos^3\beta$ function to the ADMR results shown in (b). (d) $B$ dependence of the first- and third-order SHAHE. The blue (red) plots show the first-order (third-order) SHAHE, $S^{1st}_{SHAHE}$ ($S^{3rd}_{SHAHE}$), obtained from the ADMR. The deep blue curve indicates $S_{SHAHE}$ obtained from the FDMR results.

**APPENDIX F: MOLECULAR FIELD APPROXIMATION FOR MAGNETIZATION OF GGG**

GGG is an ideal Curie-paramagnet with a weak exchange interaction between spins of neighboring Gd ions. Using the molecular field approximation, the thermal average of spin $\langle m \rangle$ is calculated by the self-consistent equation [48]

$$\langle m \rangle = -S B_S (S C_1 B_{\text{eff}} / T), \quad (F1)$$

where $S$ is the electron spin angular momentum of Gd ions, $B_s(x)$ is the Brillouin function of spin $S$ as a function of $x$, $C_1 = g\mu_B/k_B$, $g$ is the $g$-factor, $\mu_B$ is the Bohr magneton, $B_{\text{eff}} = B + N_{\text{PI}} \langle m \rangle J_{\text{ex}} / g\mu_B$ is the effective field including the applied magnetic field $B$ and the Weiss molecular fields, $k_B$ is the Boltzmann constant, $T$ is the temperature, $N_{\text{PI}}$ is the number of the nearest neighbor of the interfacial magnetic moments and $J_{\text{ex}}$ is the strength of the antiferromagnetic exchange interaction among Gd ions.

We used the effective (renormalized) Curie-Weiss temperature $\Theta_{\text{CW}}^{\text{eff}}$ for taking all the correlation effects on a Gd ion into account, which gives the effective field as $B_{\text{eff}} = BT/(T - \Theta_{\text{CW}}^{\text{eff}})$. The $\Theta_{\text{CW}}^{\text{eff}}$ should recover the bare Curie-Weiss temperature $\Theta_{\text{CW}}$ in the limit $B \to 0$ and $3T\Theta_{\text{CW}}/[C_1(S+1)B]$ in the limit $B \to \infty$, respectively. For practical purposes it is convenient to match these limiting cases into a crossover function for the effective Curie-Weiss temperature:

$$\Theta_{\text{CW}}^{\text{eff}}(B) = \frac{3\Theta_{\text{CW}}}{S+1} \frac{B_S(S\xi)}{\xi} \approx \begin{cases} \Theta_{\text{CW}} & (g\mu_B B/k_B T \ll 1) \\ \frac{3T}{C_1(S+1)B} \Theta_{\text{CW}} & (g\mu_B B/k_B T \gg 1) \end{cases}, \quad (F2)$$

with the ansatz $\xi = -a_0 + a_1|B| + \sqrt{a_0^2 + (a_2 B)^2}$, where $a_0 = -3\Theta_{\text{CW}}/(S+1)T$, $a_1 = C_1/(T-\Theta_{\text{CW}})$, and $a_2 = C_1/T(1-T/\Theta_{\text{CW}})$.

Figure 11 shows the plots of Eq. (F1) solved self-consistently (red line) and using with $\Theta_{\text{CW}}^{\text{eff}}$ given as Eq. (F2) (the blue line). We find a good agreement between both curves, justifying the use of $\Theta_{\text{CW}}^{\text{eff}}$. In the following discussion and the main text, the approximate form of $B_{\text{eff}} = BT/(T - \Theta_{\text{CW}}^{\text{eff}})$ is used to analyze the data.

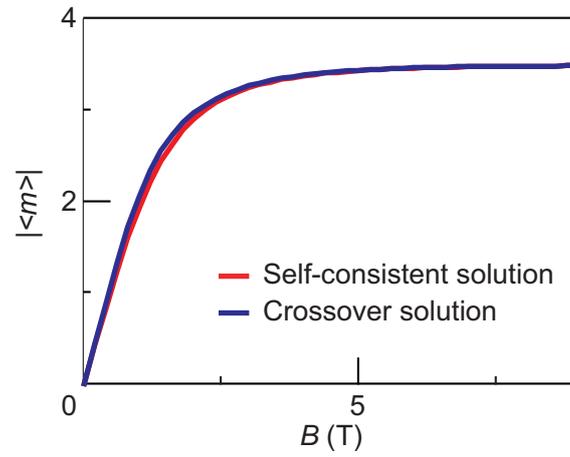

FIG. 11. *B* dependence of |<*m*>|. The red and blue lines represent the self-consistent solution and the approximation with the effective Curie-Weiss temperature solution for Eq. (F2), respectively.

**APPENDIX G: THEORY OF PARAMAGNETIC SMR**

Chen et al. formulated the spin Hall magnetoresistance (SMR) and spin Hall anomalous Hall effect (SHAHE) in a normal metal (NM)/ferromagnetic insulator (FM) bilayer system in Ref. 12. They considered the NM/FM structure shown in Fig. 12(a) and solved a spin diffusion equation with a boundary condition which describes spin transfer between a conduction electron in NM and magnetization in FM at the interface. The longitudinal ($\rho_L$) and transverse resistivity ($\rho_T$) of NM is given as

$$\rho_L \approx \rho_D + \Delta\rho_0 + \Delta\rho_1(1 - n_y^2), \tag{G1}$$

$$\rho_T \approx \Delta\rho_1 n_x n_y + \Delta\rho_2 n_z, \tag{G2}$$

with

$$\frac{\Delta\rho_0}{\rho_D} = -\theta_{SH}^2 \frac{2\lambda_{NM}}{d_{NM}} \tanh\frac{d_{NM}}{2\lambda_{NM}}, \tag{G3}$$

$$\frac{\Delta\rho_1}{\rho_D} = \theta_{SH}^2 \frac{\lambda_{NM}}{d_{NM}} \operatorname{Re} \frac{2\lambda_{NM} G_{\uparrow\downarrow} \tanh^2\frac{d_{NM}}{2\lambda_{NM}}}{\sigma_D + 2\lambda_{NM} G_{\uparrow\downarrow} \coth\frac{d_{NM}}{\lambda_{NM}}}, \tag{G4}$$

$$\frac{\Delta\rho_2}{\rho_D} = -\theta_{SH}^2 \frac{\lambda_{NM}}{d_{NM}} \operatorname{Im} \frac{2\lambda_{NM} G_{\uparrow\downarrow} \tanh^2\frac{d_{NM}}{2\lambda_{NM}}}{\sigma_D + 2\lambda_{NM} G_{\uparrow\downarrow} \coth\frac{d_{NM}}{\lambda_{NM}}}, \tag{G5}$$

where $\rho_D = 1/\sigma_D$ is the intrinsic electric resistivity of NM, $\mathbf{n} = (n_x, n_y, n_z)$ is the unit vector of the magnetization of FI, $\theta_{SH}$ is the spin Hall angle of NM, $\lambda_{NM}$ is the spin diffusion length of NM, $d_{NM}$ is the thickness of NM, and $G_{\uparrow\downarrow} = G_r + iG_i$ is the complex spin mixing conductance at the NM/FM interface. This formulation [G1 and G2] succeeded in describing the experimental observation of SMR ($\Delta\rho_1/\rho_D$) and SHAHE ($\Delta\rho_2/\rho_D$) in NM/FM systems. They modeled the coupling between conduction electron spin in NM and macroscopic magnetization of FM, which is assumed to be independent of $B$.

In the following sections, we model the paramagnetic spin Hall magnetoresistance [26]. We consider spin transfer at a paramagnetic insulator(PI)/NM interface via the interfacial exchange interaction between a conduction electron spin in NM and localized spin (not magnetization) in PI. As a result of the interaction, the spin relaxation time of the conduction electron in NM becomes anisotropic, giving rise to the SMR and SHAHE as discussed below. This approach clarifies the relation between the spin conductance and anisotropic spin relaxation in NM, which is crucial to understand the paramagnetic SMR.

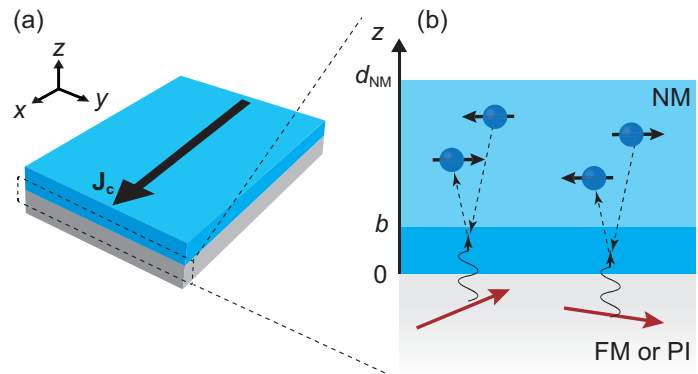

FIG. 12. (a) A schematic illustration of the NM/PI or NM/FM sample structure. We apply a charge current, $J_c$, in the $x$ direction. (b) A schematic side view of the interface. Conduction electron spins are coupled to spins in PI in the deep blue region ($0 < z < b$) with the thickness of $b$. $d_{NM}$ is the thickness of NM.

**APPENDIX H: CORRECTION FROM EFFECTIVE SPIN CONDUCTANCE AT THE INTERFACE**

First, we describe SMR with the boundary condition including the effective (longitudinal) spin conductance $G_s$. $G_s$ characterizes the spin-flip process at the interface, which is neglected in a conventional NM/FM interface [12]. For the sake of completeness, we keep $G_s$ in the boundary condition describing the spin current at the interface,

$$-e\mathbf{J}_s = G_s \boldsymbol{\mu}_s + G_r \mathbf{n} \times (\mathbf{n} \times \boldsymbol{\mu}_s) + G_i \mathbf{n} \times \boldsymbol{\mu}_s, \tag{H1}$$

where $e$ is the elementary charge, $\mathbf{J}_s$ is the spin current vector, $\boldsymbol{\mu}_s$ is the spin accumulation vector at the NM/PI interface, and $\mathbf{n} = \mathbf{B}/B$ is the unit vector of the applied magnetic field $\mathbf{B}$. By solving a one-dimensional spin diffusion equation in the $z$ direction with the boundary condition [Eq. (H1)] at $z = 0$ and the zero spin current at $z = d_{NM}$, we obtain the same form of the longitudinal and transverse resistivity with Eq. (G1) and (G2) but the different expressions of $\Delta\rho_0/\rho_D$, $\Delta\rho_1/\rho_D$, and $\Delta\rho_2/\rho_D$ as,

$$\frac{\Delta\rho_0}{\rho_D} = 2\theta_{SH}^2 \left[1 - \frac{\lambda_{NM}}{d_{NM}} \tanh\left(\frac{d_{NM}}{2\lambda_{NM}}\right) \mathcal{R}(G_s)\right], \tag{H2}$$

$$\frac{\Delta\rho_1}{\rho_D} = \theta_{SH}^2 \frac{2\lambda_{NM}}{d_{NM}} \tanh\left(\frac{d_{NM}}{2\lambda_{NM}}\right) [\mathcal{R}(G_s) - \text{Re}[\mathcal{R}(G_s - G_{\uparrow\downarrow})]]$$

$$= D_1[\mathcal{R}(G_s) - \text{Re}[\mathcal{R}(G_s - G_{\uparrow\downarrow})]], \tag{H3}$$

$$\frac{\Delta\rho_2}{\rho_D} = \theta_{SH}^2 \frac{2\lambda_{NM}}{d_{NM}} \tanh\left(\frac{d_{NM}}{2\lambda_{NM}}\right) \text{Im}[\mathcal{R}(G_s - G_{\uparrow\downarrow})] = D_1 \text{Im}[\mathcal{R}(G_s - G_{\uparrow\downarrow})], \tag{H4}$$

with

$$\mathcal{R}(x) = \frac{1 - x\rho_D\lambda_{NM}\coth(d_{NM}/2\lambda_{NM})}{1 - 2x\rho_D\lambda_{NM}\coth(d_{NM}/\lambda_{NM})} = \frac{1 - D_2 x}{1 - D_3 x}, \tag{H5}$$

where $D_1 = \theta_{SH}^2 (2\lambda_{NM}/d_{NM})\tanh(\lambda_{NM}/2d_{NM})$, $D_2 = \rho_D\lambda_{NM}\coth(d_{NM}/2\lambda_{NM})$, and $D_3 = 2\rho_D\lambda_{NM}\coth(d_{NM}/\lambda_{NM})$ are constants, which can be calculated using material parameters shown in the later.

Equations (H2)-(H4) describe the correction from $G_s$, and they are reduced to Eqs. (G3)-(G5) when $G_s = 0$, which corresponds to the same boundary condition used by Chen et al. [12].

**APPENDIX I: ANISOTOROPIC SPIN RELAXATION DUE TO INTERFACIAL SPIN EXCHANGE INTERACTION**

Next, we explain the relation between the interface spin conductances and the spin relaxation times calculated for the conduction electron in the NM in close vicinity of the NM/PI interface. We model the interface by introducing an auxiliary intermixing layer between the NM and the PI with the thickness $b$ and taking a $b \to 0$ limit for calculation [26,49] of $G_{\uparrow\downarrow}$ and $G_s$. In such layer, dark blue region in Fig. 12(b), conducting electrons couple to the localized spins in the PI via an interfacial exchange interaction. This interaction is described by the effective Hamiltonian:

$$\mathcal{H}_{\text{int}} = -\mathcal{J}_{\text{int}} \sum_i \mathbf{S}_i \cdot \mathbf{s}(\mathbf{r}_i), \tag{I1}$$

where $\mathcal{J}_{\text{int}}$ is the coupling constant, $\mathbf{S}_i$ is the localized spin operator and $\mathbf{s}(\mathbf{r}_i)$ is the spin density of conduction electrons at position $\mathbf{r}_i$. The continuity equation for the spin current in the interaction region $0 < z < b$ reads

$$\partial_t \mu_s^\alpha - \frac{1}{e\nu_F} \partial_i j_{s,i}^\alpha - \omega_L(\mathbf{r}) \epsilon_{\alpha\beta\gamma} n_\beta \mu_s^\gamma = -\Gamma_{\alpha\gamma} \mu_s^\gamma, \tag{I2}$$

where the superscript Greek indices denote spin projections ($\alpha, \beta, \gamma = x, y, z$), subscript Latin ones (such as $i = x, y, z$ but not the descriptor index "s") denote current directions, $\mu_s^\alpha$ is the spin accumulation polarized in the $\alpha$ direction, $\nu_F$ is the density of states at the Fermi level, $j_{s,i}^\alpha$ is the spin current flowing in the $i$ direction and spin-polarized in the $\alpha$ direction, $\omega_L = \omega_B - \delta_b(z) \langle \hat{S}_\parallel \rangle (n_{\text{PI}} \mathcal{J}_{\text{int}}/\hbar)$ is the effective (renormalized) Larmor frequency, $\omega_B = g\mu_B B/\hbar$ is the bare Larmor frequency, $g$ is the $g$-factor, $\mu_B$ is the Bohr magneton, $\hbar$ is the Dirac constant, $\delta_b(z) = 1/b$ $(0 < z < b)$, $0$ $(z > b)$, $\langle \hat{S}_\parallel \rangle$ is the expectation of spin parallel to $\mathbf{B}$, $n_{\text{PI}}$ is the number of localized spins per unit area at the surface of the PI, and $\Gamma_{\alpha\gamma}$ is the spin relaxation tensor. Here, $\omega_L$ is renormalized by the interfacial exchange fields [18] due to $\langle \hat{S}_\parallel \rangle$. For the case with uniaxial symmetry set by $\mathbf{B}$, $\Gamma_{\alpha\gamma}$ has the general form:

$$\Gamma_{\alpha\gamma}(r) = \frac{\delta_{\alpha\gamma}}{\tau_s} + \delta_b(z) \left[ \frac{\delta_{\alpha\gamma}}{\tau_\perp} + \left( \frac{1}{\tau_\parallel} - \frac{1}{\tau_\perp} \right) n_\alpha n_\gamma \right], \tag{I3}$$

where $\tau_s$ is the isotropic part of spin relaxation time induced by the spin-orbit coupling or magnetic impurities in NM, and $\tau_\perp$ and $\tau_\parallel$ is the transverse and longitudinal spin relaxation time per unit of thickness in the interaction region, respectively.

By combining Eq. (I2) with Eq. (I3), we obtain the spin current in the region where the exchange interaction takes place [dark blue region in Fig. 12(b)] as

$$-\frac{1}{e\nu_F} j_{s,z}^\alpha \Big|_{z=0}^{z=b} = b\omega_L \epsilon_{\alpha\beta\gamma} n_\beta \mu_s^\gamma - \left( \frac{b}{\tau_s} + \frac{1}{\tau_\perp} \right) \mu_s^\alpha - \left( \frac{1}{\tau_\parallel} - \frac{1}{\tau_\perp} \right) n_\alpha (\mathbf{n} \cdot \boldsymbol{\mu}_s). \tag{I4}$$

We take a $b \to 0$ limit to describe the interface spin current, where $\delta_b(z)$ becomes a delta function. We compare Eqs. (H1) and (I4) and obtain the relation between the interfacial spin conductances and anisotropic spin relaxation times,

$$G_r = e^2 \nu_F \left( \frac{1}{\tau_\perp} - \frac{1}{\tau_\parallel} \right), \tag{I5}$$

$$G_i = -\frac{e^2}{\hbar} n_{PI} \mathcal{J}_{int} \langle \hat{S}_\parallel \rangle, \tag{I6}$$

$$G_s = -e^2 \nu_F \frac{1}{\tau_\parallel}, \tag{I7}$$

where $\nu_F \mathcal{J}_{int} = J_{int}$ is the dimensionless interfacial exchange interaction.

In order to determine the anisotropic spin relaxation times $\tau_\perp$ and $\tau_\parallel$, we use the Born-Markov approximation [50] and obtain [26]

$$\frac{1}{\tau_\parallel} = \frac{2\pi}{\hbar} \frac{n_{PI} J_{int}^2}{\nu_F} \xi n_B(\xi)[1 + n_B(\xi)] |\langle \hat{S}_\parallel \rangle|, \tag{I8}$$

$$\frac{1}{\tau_\perp} = \frac{1}{2\tau_\parallel} + \frac{\pi}{\hbar} \frac{n_{PI} J_{int}^2}{\nu_F} \langle \hat{S}_\parallel^2 \rangle, \tag{I9}$$

where $n_B(\xi) = 1/(e^\xi - 1)$ is the Bose-Einstein distribution as a function of $\xi = g\mu_B B_{eff}/k_B T = C_1 B_{eff}/T$, and $B_{eff}$ is the effective magnetic field of the PI (see Appendix F). In a paramagnetic phase, $|\langle \hat{S}_\parallel \rangle|$ and $\langle \hat{S}_\parallel^2 \rangle$ in Eqs. (I8) and (I9) can be determined as [26],

$$\langle \hat{S}_\parallel \rangle = -S B_S(S\xi), \tag{I10}$$

$$\langle \hat{S}_\parallel^2 \rangle = S(S+1) - \coth(\xi/2) S B_S(S\xi), \tag{I11}$$

where $S$ is the spin of $Gd^{3+}$ in GGG.

Importantly, the difference between the longitudinal and transverse spin relaxation times appears only for a finite $|\langle \hat{S}_\parallel \rangle|$. At the zero magnetic field, Eqs. (I8) and (I9) become the same, then

$$\frac{1}{\tau_\parallel} = \frac{1}{\tau_\perp} = \frac{2\pi}{3\hbar} \frac{n_{PI} J_{int}^2}{\nu_F} S(S+1). \tag{I12}$$

Substituting the above results into Eqs. (I5)-(I7), we obtain the magnetic field dependence of the spin conductances at a NM/PI interface:

$$G_r = \frac{\pi}{\hbar} n_{PI} (e \mathcal{J}_{int})^2 \left\{ S(S+1) - \left[ \coth(\xi/2) + \frac{\xi}{4\sinh^2(\xi/2)} \right] S B_S(S\xi) \right\}$$

$$= A_1 \left\{ S(S+1) - \left[ \coth(\xi/2) + \frac{\xi}{4\sinh^2(\xi)} \right] S B_S(S\xi) \right\} \tag{I13}$$

$$G_i = \frac{e^2}{\hbar} n_{\text{PI}} J_{\text{int}} S B_S(S\xi) = A_2 S B_S(S\xi), \tag{I14}$$

$$G_s = -\frac{\pi}{\hbar} n_{\text{PI}} (eJ_{\text{int}})^2 \frac{\xi}{2\sinh^2(\xi/2)} S B_S(S\xi) = -A_1 \frac{\xi}{2\sinh^2(\xi/2)} S B_S(S\xi), \tag{I15}$$

where $A_1 = (\pi/\hbar) n_{\text{PI}} (eJ_{\text{int}})^2$ and $A_2 = (\pi/\hbar) n_{\text{PI}} (e^2 J_{\text{int}})$.

From the fitting to the experimental data we determine three free parameters, $\Theta_{\text{CW}}$, $n_{\text{PI}}$, and $J_{\text{int}}$. $\Theta_{\text{CW}}$ is obtained from the effective Curie-Weiss temperature $\Theta_{\text{CW}}^{\text{eff}}$, while $n_{\text{PI}}$ and $J_{\text{int}}$ from the values of $A_1$ and $A_2$. The Pt resistivity of $3.4\times10^{-7}$ $\Omega$m predicts the following parameters [51]: $\theta_{\text{SH}} = 0.104$, $\lambda_{\text{Pt}} = 2$ nm. We fit Eqs. (H3) and (H4) to the measured transverse FDMR results of SMR and SHAHE and obtain $n_{\text{PI}} = 6.94\times10^{16}$ atom/m$^2$, $J_{\text{int}} = -0.13$, and $\Theta_{\text{CW}} = -1.26$ K. The lower Curie-Weiss temperature than the bulk value of - 2 K suggests the local moments on the surface of GGG have a lower coordination number than the bulk. This is consistent with the lower concentration of spins at the GGG/Pt interface than in bulk GGG, $n_{\text{PI}} = 6.9\times10^{18}$ atom/m$^2$. Taking a typical value of density of state for a metal [52], $v_F \approx 3.5\times10^{28}$ m$^{-3}$eV$^{-1}$, the corresponding antiferromagnetic $s$-$f$ exchange coupling is about -2 meV, which is similar to the obtained for a Pt/EuS interface [18,44], -3~-4 meV. Our theoretical framework describing the spin transport at the NM/PI interface and the analysis of SMR established here can be applied to results in other magnets including para-, ferri-, ferro-, and antiferromagnets [26].

**APPENDIX J: TEMPERATURE DEPENDENCE OF THE EXPERIMENTAL AND THEORETICAL RESULTS**

Here we discuss the $T$ dependence of SMR in our system by comparing the experimental and theoretical results. Figure 13(a) shows the detailed $T$ dependence of $S_{\text{SMR}}$. If one assumes that in Eqs. (H2)-(H5) the charge and spin-transport parameters of the Pt film (such as conductivity, spin diffusion length and spin Hall angle) are *T-independent*, our theoretical model with the used parameters predicts a $1/T^2$ dependence for high temperatures, $T \gg S g \mu_B B$ (which in our case corresponds to $\sim 15$ K). However, a direct fit to the measurement shows a slower power-law decay, $\sim 1/T^{0.5}$ in the discussed $T$ range.

First of all, to reconcile the experiment and theory, we take the *T-dependence* of the parameters into account. From the $T$ dependence of the Pt resistivity (see Fig. 6), we extract their values at each $T$ based on the scaling law in Ref. 51. After substituting these values into Eqs. (H2)-(H5), we obtained a good agreement between the experiment and theory for the power-law decay of SMR with $T$. However, the amplitude of the measured signal at large $T >$ 50 K is larger than the one obtained from the theoretical model.

According to the theory, the paramagnetic SMR dominates $S_{\text{SMR}}$ below 50 K [the shaded region in Fig. 13(a)] and a strong reduction of $S_{\text{SMR}}$ at high $T$ is expected. In contrast, a sizable signal is observed up to 200 K with a weaker $T$ dependence. Most likely, this spurious signal stems from a spin subsystem with a much broader $T$ dependence than the paramagnetic SMR. To confirm this, we subtract this "background" signal from $S_{\text{SMR}}$ below 50 K. Figure 14(a) shows the $T$ dependence of $U_{\text{SMR}}(T) = S_{\text{SMR}}(T) - S_{\text{SMR}}(T = 50$ K$)$ as blue solid plots. Interestingly, after this "background" subtraction, the amplitude of the SMR signal is in very good agreement with the theory. This indicates that at large $T$ when the paramagnetic SMR becomes negligible, we clearly detect another SMR-like magnetoresistance with a weaker $T$ dependence.

This finding, together with the good agreement between the model and experiment on the $B$ dependence, confirms that at low $T$ ($< 50$ K) the observed magnetoresistance is attributed to the paramagnetic SMR. Figures 14(b), (c), and (d) show the best fits of our model to $U_{\text{SMR}}(B)$ and $S_{\text{SHAHE}}(B)$ at $T = 2$ K, 5 K, and 10 K, respectively. $U_{\text{SMR}}(B)$ is obtained by subtracting $S_{\text{SMR}}(B)$ at 50 K presented in Fig. 13(b). The model again reproduces $U_{\text{SMR}}(B)$ at all $T$. In Table 1, we summarize the value of the parameters obtained from the best fitting to $U_{\text{SMR}}$ and $S_{\text{SHAHE}}$. Our model can explain the $B$ dependence of $U_{\text{SMR}}$ at different $T$ in the low $T$ regime ($T < S g \mu_B B \sim 15$ K) with similar parameters, which indicates that the origin of $U_{\text{SMR}}$ is the paramagnetic SMR. The long tail observed at high $T$ is clearly not the effect we are focusing on and, as demonstrated, a simple subtraction of such background can reveal the paramagnetic SMR.

From our transport experiments, we cannot infer the origin of the "background" signal. This requires an investigation which is beyond the scope of this manuscript. Nevertheless, the data suggest the existence of a paramagnetic subsystem with broader $T$-dependence. Plausibly, it could be composed of a small amount of Gd atoms absorbed into Pt during its deposition on the GGG surface by sputtering. These Gd atoms couple much stronger to electrons in Pt than the ones at the interface, and their $T$ dependence is expected to be broader than Gd ions in GGG with different characteristic scale $Sg\mu_B B$. To identify the absorbed Gd atoms, we performed scanning transmission electron microscopy (STEM) and energy-dispersive X-ray spectroscopy (EDX). Figure 15(a) shows the STEM image across the Pt/GGG interface. While the sample has a reasonably good interface, we found an amorphous layer with a very small thickness of about 0.5 nm, which was created by the Pt sputtering. Importantly, the EDX profile [Fig. 15(b)] shows the amorphous layer consists of Pt, Ga and Gd atoms, indicating a small amount of Gd atoms are absorbed into Pt. Besides, at the amorphous layer, the Ga atoms show broader distribution than Gd, and it makes the majority of Gd atoms at the surface separated from Pt. These findings support the existence of a paramagnetic subsystem. Other possible factors that, combined with the above effect, may contribute to the observed background signal is the renormalization of the effective $s$-$d$ coupling ($J_{int}$) either by the Kondo effect that may be important in Pt [53], or phonons. Any of these effects may have a rather different $T$ dependence with respect to SMR. Their studies are beyond the scope of our work, but definitely, it may be interesting to explore them using SMR in future experiments.

It is important to remark that the above discussion does not change substantially the results and discussion in the main manuscript. The subtracted background signal, $S_{SMR}(T = 50\,\text{K})$, amounts only to about 19% of $S_{SMR}(T = 2\,\text{K})$, indicating that the low-temperature $S_{SMR}$ is dominated by the paramagnetic SMR. In addition, the obtained parameters at 2 K from $S_{SMR}(B)$ is consistent with that from $U_{SMR}(B, T)$. Therefore, we can confirm that our manuscript is adequately based on the results of the paramagnetic SMR.

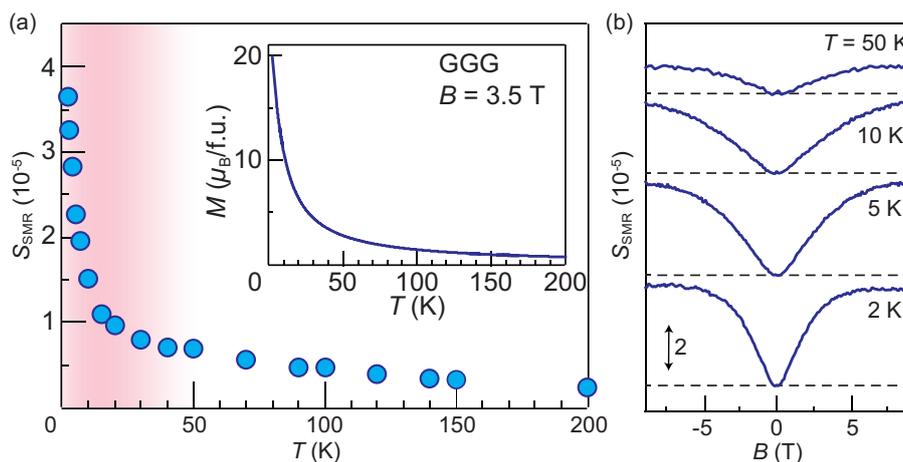

FIG. 13. (a) $T$ dependence of paramagnetic SMR and magnetization of GGG. $S_{SMR}$ is estimated by fitting a $S_{SMR}\sin(\alpha)\cos(\alpha)$ function to the transverse ADMR results at $|B| = 3.5$ T at various $T$. The inset shows the $T$ dependence of $M$ of GGG at $|B| = 3.5$ T. The paramagnetic SMR dominates the signal in the shaded $T$ region. (b) FDMR result of $S_{SMR}$ at selected $T$. The solid blue curves represent $S_{SMR}^{FDMR} = \Delta\rho_T(45°) - \Delta\rho_T(135°)$.

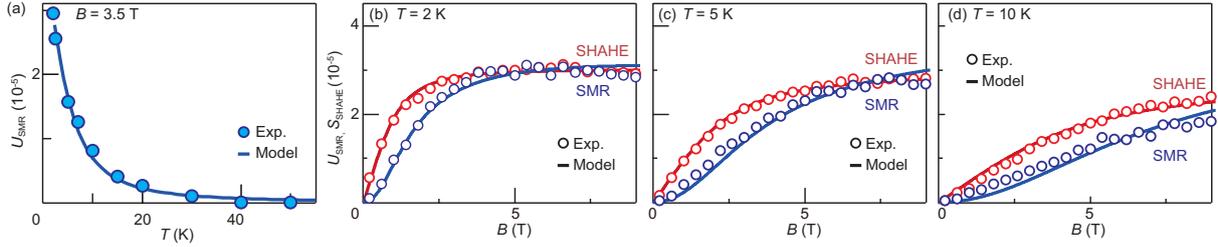

FIG. 14. (a) $T$ dependence of SMR with the model result at $|B| = 3.5$ T. We subtract $S_{SMR}(T = 50 K)$ from $S_{SMR}(T)$ and obtain $U_{SMR}(T) = S_{SMR}(T) - S_{SMR}(T = 50 K)$. The solid curve shows the best fitting of Eq. (5) as a function of $T$. (b)-(d) $B$ dependence of $U_{SMR}$, and $S_{SHAHE}$ at (b) 2K, (c) 5 K, and (d) 10 K with the model results. We fit Eqs (5) and (6) to each experimental results shown as the unfiled plots.

TABLE 1. Parameter values for best fitting.

|  | $U_{SMR}(T)$ | $U_{SMR}(B)$ (2 K) | $U_{SMR}(B)$ (5 K) | $U_{SMR}(B)$ (10 K) | $S_{SMR}(B)$ (2 K) |
|---|---|---|---|---|---|
| $n_{PI}$ (Gd atm/m$^2$) | 6.94×10$^{16}$ | 9.12×10$^{16}$ | 7.68×10$^{16}$ | 6.47×10$^{16}$ | 6.89×10$^{16}$ |
| $\theta_{CW}$ (K) | -1.27 | -0.77 | -1.00 | -1.27 | -1.26 |
| $J_{int}$ | -0.25 | -0.09 | -0.10 | -0.12 | -0.13 |

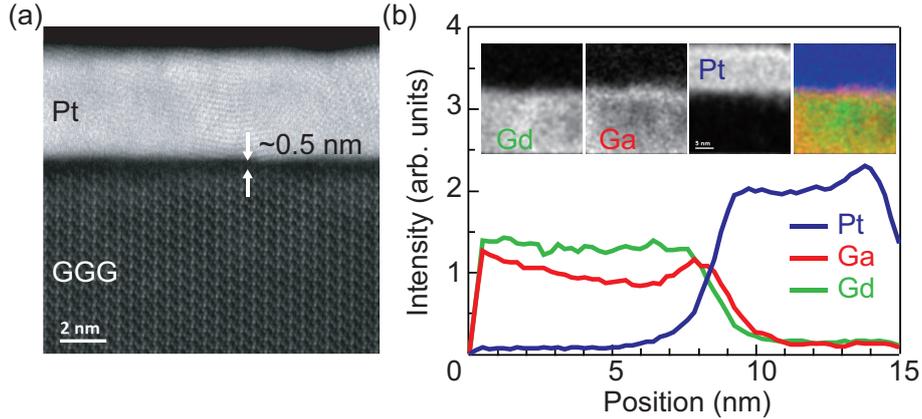

FIG. 15. (a) STEM image of the Pt/GGG junction. (b) The spatial distribution of elements along the Pt/GGG interface probed by EDX. The blue, red, and green lines indicate the intensity of Pt, Ga, and Gd atoms, respectively. The inset is the image of the distribution of Pt, Ga, and Gd atoms, and the color-coded image at the interface.